\documentclass[3p,twocolumn]{elsarticle}

\usepackage{dcolumn}
\usepackage{bm}
\usepackage[version=3]{mhchem} 

\usepackage{amssymb}
\usepackage{amsthm}
\usepackage{amsmath}
\usepackage{hyperref}
\usepackage{cases}
\hypersetup{pdfborder={0 0 0}, colorlinks=true, citecolor=blue, linkcolor=blue, urlcolor=blue}
\usepackage{natbib}
\RequirePackage{tikz}
\RequirePackage{pgfplots}
\usepgfplotslibrary{fillbetween}
\usetikzlibrary{datavisualization.formats.functions} 
\usetikzlibrary{decorations.text}
\usepackage{wrapfig}
\usepackage{pbox}

\usepackage{subcaption}
\usepackage{graphicx}
\usepackage{placeins}
 
\journal{Journal of Engergy Science and Technology}
\begin{document}

\begin{frontmatter}

\title{Simple Model for Estimation of the Influence of Velocity on Advancing Dynamic Contact Angles}
\author{Pooyan.~Heravi}
\address{Department of Power Mechanical Engineering, National Tsing Hua University}
\author{Zung-Hang~Wei\corref{cor1}}
\ead{wei@pme.nthu.edu.tw}
\address{Department of Power Mechanical Engineering, National Tsing Hua University}
\address{Institute of NanoEngineering and MicroSystems, National Tsing Hua University}
\author{Farschad.~ Torabi\corref{cor1}}
\ead{ftorabi@kntu.ac.ir}
\address{Department of Energy Systems, K.N. Toosi University of Technology}

\cortext[cor1]{Corresponding author}

\begin{abstract}
Given the importance of dynamic contact angle, its numerous applications and the complexity and difficulty of use of available approaches, here we present a new simple semi-empirical models for estimation of dynamic contact angle's dependence on the wetting line velocity. These models should be applicable to any geometry, a very large range of capillary numbers and static contact angles and all solid-liquid-gas systems without requiring further experiments. Two simple equations are intuitively derived from the most promising theoretical dynamic contact angle models, the hydrodynamic and the molecular-kinetic models. Then the models, along with the basic form of the Hoffman model, are fitted to a large pool of data. The data are extracted from numerous studies and cover over 5 decades of capillary number, include static contact angles up to the superhydrophobic region and comprise of various geometries. The resulting models are compared to each other where the hydrodynamic model's predictions are found to be superior to the other two models by all statistical measures. Then, noting that the molecular forces become dominant at lower capillary numbers we separate the data into low and high capillary regions and repeat the process. There was only a minuscule difference between the results obtained for general models and high capillary models ($\textrm{Ca}>10^{-4}$) but the empirical approach resulted in the most accurate model at low capillary numbers ($\textrm{Ca}<10^{-4}$). 
\end{abstract}

\begin{keyword}

Interfacial Flow  \sep Dynamical Wetting \sep Dynamic Contact Angle Prediction

\end{keyword}
\end{frontmatter}

\section{Introduction}

Dynamic contact angles have been a topic of interest and debate for decades, yet despite significant theoretical advances we still have limited understanding of the phenomena that governs them. The bulk of scientific research in recent years have been dedicated to proving or disproving proposed theoretical approaches but none of these efforts have produced a definitive answer and there isn't even a hint of consensus among researchers as to which model would ultimately provide adequate prediction capabilities~\cite{snoeijer2013moving}. Furthermore, these theoretical approaches also require experimental parameters that are not known \textit{a priori} (at least for now). These parameters are obtained by fitting the theoretical models to the dynamic contact angle profile for every particular set of solid, liquid and gas system. Therefore these models have yet to become capable of predicting dynamic contact angles.

The latest efforts to find a model that can be readily used to predict contact angle's dependence on velocity date back to the 20th century when in 1992 Seebergh et al.~\cite{seebergh1992dynamic} experimentally investigated the relationship and produced models by fitting various functions to the experimental data. But this useful approach has been largely abandoned in favor of the promising and more recently proposed theoretical models, which as previously stated have not yet resulted in a concrete and usable dynamic contact angle model.

All the while, scientists and engineers are facing problems where dynamic contact angles play a major role. Dynamic contact angles appear in a myriad of natural phenomena and industrial applications, including droplet dynamics, coating flows, microfluidic applications and two-phase flows in porous media and capillary tubes~\cite{yarin2006drop,kistler1997liquid,darhuber2005principles,adler1988multiphase,heravi2016mathematical}. Moreover, the current shift to small-scale devices means that capillary forces and contact angles will become even more important as we move forward. To further their efforts, scientists that work on these subjects require models to predict dynamic contact angles, therefore they have no choice but to use models that were obtained more than a decade ago and have not been updated as more experimental data has been published.

Furthermore, there have been various claims, usually made by those who have proposed the models in the first place, about capabilities of a model and its superiority over another. Yet, some studies have shown that limited experimental data sets could actually be more or less accounted for by any one of the models~\cite{seveno2009dynamics,ranabothu2005dynamic}.

In the authors' opinion, the shortcomings of previous studies might stem from the relatively limited data pool used in each study. Therefore in this study results from various sources have been extracted in order to compile an adequate data pool. The data cover 5 decades of capillary numbers~($\textrm{Ca}=\mu u/\sigma$) and include static contact angles up to superhydrophobic angles.

The present study attempts to provide a simple intuitive model that is capable of predicting the advancing dynamic contact angle with acceptable accuracy over a wide-range of flow conditions and static contact angles for various moving dynamical wetting transition problems. To this end we will present a brief description of the most commonly used dynamic contact angle models. Then we will derive a general form for each model and find the best possible fit based on the general forms. Finally, we will compare prediction capabilities of these models in order to find the most accurate one.

\section{Dynamic Contact Angle Models}
In this section we will first briefly describe the dynamic contact angle phenomena. Afterwards, we will discuss the hydrodynamic, molecular-kinetic and empirical models and attempt to predict experimentally measured dynamic contact angle values by using each model. Then, we will manipulate the models in order to derive a form that relates a function of the contact angle to a function of velocity or capillary number. We will exclude some rather well known variations of these models as they are much more complex and would not lend themselves well to our approach.

\begin{figure}
    \centering
    \def\svgwidth{\columnwidth}
    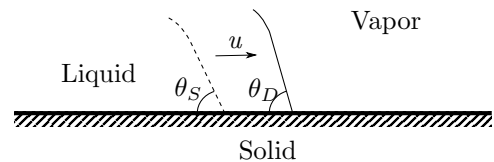
    \caption{Schematic illustration of an advancing contact line and the dynamic contact angle}
\label{schematic}
\end{figure}

Figure~\ref{schematic} is an illustration of the problem at hand. Here $u$ represents contact line velocity while $\theta_{\textrm{S}}$ and $\theta_{\textrm{D}}$ denote static and dynamic contact angles respectively. When the contact line is stationary the contact angle equals $\theta_{\textrm{S}}$ (assuming zero contact angle hysteresis), but contact angle changes as soon as the contact line begins moving. The exact value of the change in contact angle depends on various parameters, most notably the contact line velocity, surface tension and viscosity. It should be noted that we are only addressing the problem of an advancing dynamic contact angle and do not engage more complicated problems, e.g. the receding contact line, contact line over a heterogeneous surface or Marangoni effects.
\subsection{Hydrodynamic Model}
As one might expect, the first theoretical modeling attempts utilized hydrodynamics in order to formulate a model for the dynamic contact angle. Yet, the dynamic contact angle phenomenon is a three-phase problem where the interactions close to the three-phase line play a major role and that's where problems arise. The contradiction of a moving contact line and the traditional approach of assuming zero velocity at walls result in unphysical tensions very close to a moving contact line. Cox-Voinov law is an attempt to solve this problem by using an approximation of the Stoke's equations known as the lubrication approximation. Below is a simplification of this law derived by Blake and Ruschak~\cite{bonn2009wetting} for contact angles smaller than $\frac{3\pi}{4}$:
\begin{equation}
\theta^{3}_{\textrm{D}}-\theta^{3}_{\textrm{S}}=9\frac{\mu u}{\sigma} \textrm{ln}\Big(\frac{L}{L_m}\Big)=\Phi(\textrm{Ca})
\label{hydro}
\end{equation}
where $\mu$ is viscosity, $\sigma$ is surface tension, $\Phi(\textrm{Ca})$ represents a function of $\textrm{Ca}$ number and $L$ and $L_m$ are two parameters required for Cox and Voinov's approximation. $L$ is a macroscopic characteristic length scale, e.g. a droplet's diameter, while $L_m$ is used to truncate the solution at small scales and is a a representation of the microscopic phenomena that solve the singularity at the interface. There is no consensus over how to obtain $L$ and $L_m$ theoretically and some studies have reported that the parameters assumed to be constant in the hydrodynamic model may in fact depend on flow properties such as contact line velocity~\cite{rame2004characterizing}. Therefore for our purposes of a general, versatile and simple model, we simply assume that the difference of the cubes of static and dynamic contact angles is a function of capillary number.

\subsection{Molecular-Kinetic Theory}
Another model known as molecular-kinetic theory (MKT) provides a contrasting explanation for dynamic contact angles. According to MKT dissipation at a moving contact line is dominated by the thermodynamic processes at molecular scales. Here the contact line motion is modeled by using two empirical parameters, length of molecule displacements $\lambda$ (the jump length (m)) and the frequency of the jumps from one adsorption site to another characterized by $K_0$ (Hz)~\cite{seveno2009dynamics}. 
\begin{equation}
u=2\kappa^{0}\lambda \sinh[\sigma(\cos\theta_{S}-\cos\theta_{D})\lambda^2/2k_\mathrm{B}T]
\label{MKT}
\end{equation}
where $k_B$ is the Boltzmann constant, \textit{T} is the absolute temperature and $\kappa$ is the equilibrium frequency of random molecular displacements. 

This model can be simplified by substituting the hyperbolic sine with its Taylor expansion and assuming that the argument of the hyperbolic sine is small, hence one will obtain the linearized form of MKT:
\begin{equation}
u=\kappa^0 \lambda^3 \sigma(\cos\theta_{S}-\cos\theta_{D})/k_B T
\label{linMKT}
\end{equation}
Further simplification can be achieved by solving the equation for the term in the paranthesis and replacing it with its Taylor expansion which will result in:
\begin{equation}
\theta^{2}_{\mathrm{D}}-\theta^{2}_{\mathrm{S}}=u k_B T/\kappa^0 \lambda^3 \sigma=\zeta u/\sigma
\label{MKTd}
\end{equation}
where $\zeta$ has the same unit as viscosity and can be thought of as the coefficient of friction at the contact line. Furthermore, the group on the right hand of this equation forms a dimensionless group and thinking of $\zeta$ as viscosity would imply that this dimensionless group is similar to capillary number. Therefore, we will use this form of the model as we believe it offers a an intuitive explanation. 

\subsection{Empirical Correlations}
As previously mentioned there are also a couple of empirical models, obtained more than a decade ago, that are the only viable choices for dynamic contact angle prediction at the moment. The most widely used and experimentally verified of these models are in the form of~\hyperref[corr]{equation~(\ref{corr})} which was first proposed by Jiang et al.~\cite{jiang1979correlation} by fitting the correlation on the experimental data of Hoffman~\cite{hoffman1983study}:
\begin{equation}
H=\frac{\cos\theta_\mathrm{\alpha}-\cos\theta_\mathrm{\beta}}{\cos\theta_\mathrm{\alpha}+1}=a\mathrm{Ca}^b
\label{corr}
\end{equation}
where $\alpha$ and $\beta$ are the static and dynamic contact angles respectively, and \textit{a} and \textit{b} are constants. The values of \textit{a} and \textit{b} are equal to 2.24 and 0.54 at low capillary numbers ($\mathrm{Ca}>\mathrm{Ca}_T$, where $\mathrm{Ca}_T$) and 4.47 and 0.42 at higher capillary numbers~\cite{seebergh1992dynamic}. This model is simple and suitable for a universal approach, therefore it can be used without any manipulation.

\section{Fitting Procedure}
Having the general model forms, we fitted the following models to the data, with $a$ and $b$ as fitting parameters, using non-linear least squares method:
\begin{itemize}
\item{Empirical: for this model we have retained the exact form that was used in previous studies as it is very easy to use and has proven to be versatile:\begin{equation}H=a\mathrm{Ca}^b\end{equation}}
\item{Molecular-kinetic: $\zeta$ is an unknown term in~\hyperref[MKTd]{equation~(\ref{MKTd})} and there is no reliable method to predict $\zeta$. Therefore we have treated it as a free parameter to be obtained through fitting. After experimenting with various functions, we found out that a power law function gives good results. Substituting an exponential function into the model we get:
\begin{equation} \sigma(\theta^{2}_{\mathrm{D}}-\theta^{2}_{\mathrm{S}})=a u^{b}\end{equation}}
\item{Hydrodynamic: As previously mentioned the theoretical forms of this model is also not suitable for general purpose use and requires fitting to experiments for every situation. Therefore, noting the ambiguities surrounding characteristic lengths~\cite{rame2004characterizing} and their possible dependence on velocity, we assumed that the right hand is a function of capillary number. A power law gave one of the best matches with experimental data, therefore we substituted the right hand of this model with a power law equation, arriving at:
\begin{equation}\theta^{3}_{\mathrm{D}}-\theta^{3}_{\mathrm{S}}=a\mathrm{Ca}^b\end{equation}}
 \end{itemize}
In order to evaluate the models for a wide range of hydrodynamic conditions and fluid properties, we used values from various studies. A list of the experiments and their respective flow conditions and static contact angles can be found in table~\ref{data}. To ensure the resulting models versatility, the experiments are chosen such that they would be diﬀerent from each other in many ways. The experiments include various ranges of capillary numbers, static contact angles, measurement method and geometry. For example: Hoﬀman et al~\cite{hoffman1983study} measured contact angles for the advancing menisci in a capillary tube, Seebergh et al~\cite{seebergh1992dynamic} (using microtensiometry), Kim et al~\cite{kim2015dynamic} (using optical illumination) and Schneemilch et al~\cite{schneemilch1998dynamic} measured advancing contact angles in the movement of a Wilhelmy plate through an interface, Wang et al~\cite{wang2007spreading} and Seveno et al~\cite{seveno2009dynamics} optically measured contact angles during drop spreading and Rio et al~\cite{rio2005boundary} optically measured contact angles during the sliding of a drop down a plane.

We also took three measures to control the data. First, the studies include various amounts of data, from just a few points to thousands of points, therefore we used data point weights in order to prevent over-represenation of any particular experiment due to the difference in the number of data points. Second, the data cover a wide range of static contact angles that begin with 0 and ends near superhydrophobic static contact angles. The reason for exclusion of larger contact angles is reports presented in previous experiments which have shown that superhydrophobic static contact angles behave differently and do not change with contact line velocity~\cite{kim2015dynamic}. Third, we also imposed an upper limit of 0.1 on capillary number, since higher values would contradict major assumptions used in formulation of underlying theoretical models. 

   \begin{table*}
   \centering
    \caption{Exprimental studies included in the fitting procedure.}
    \label{data}
    \setlength{\tabcolsep}{1.5pt}
   \begin{tabular}{ lrlc }
    \hline
     \multicolumn{1}{c}{Study} & \multicolumn{2}{c}{Flow Conditions}  & \quad Static contact angle\\ \hline
    Hoffman et al.~\cite{hoffman1983study} & $2\times10^{-3}$&$<\mathrm{Ca}<2\times10^{-1}$ & $0$ \vspace{1pt}\\ 
    Seebergh et al.~\cite{seebergh1992dynamic} & $10^{-7}$&$<\mathrm{Ca}<10^{-2}$ & \pbox{20cm}{\relax\ifvmode\centering\fi
    $0, 5, 12, 35, 41, 49,$ \\ $ 57, 59$}\vspace{1pt}\\
    Kim et al.~\cite{kim2015dynamic} & $3\times10^{-4}$&$<\mathrm{Ca}<10^{-1}$ & 98  \\ 
    \pbox{20cm}{\relax\ifvmode\fi
    \leftskip0ptWang et al.~\cite{wang2007spreading} \\ (Newtonian fluid only)}\newline & $10^{-5}$&$<\mathrm{Ca}<5\times10^{-4}$ & 0  \\ 
    Schneemilch et al.~\cite{schneemilch1998dynamic} & $10^{-4}$&$<\mathrm{Ca}<2\times10^{-3}$ & 42  \\ 
    Rio et al.~\cite{rio2005boundary} & $2\times10^{-4}$&$<\mathrm{Ca}<7\times10^{-3}$ & 59  \\ 
    Seveno et al.~\cite{seveno2009dynamics}   & \quad$5\times10^{-7}$&$<\mathrm{Ca}<2\times10^{-2}$ & 0, 10, 38.8 \\  \hline
    \end{tabular}
     \end{table*}
\section{Results and Discussions}
The models are fitted to the data per the descriptions in the previous section and then the following statistical measures for absolute error ($\epsilon$) and relative error ($\eta$) are calculated for each model:
\begin{align}
\overline{\epsilon}=&\frac{1}{N}\sum_1^N\theta_{est}-\theta_{exp} \\
RMSE_{\mathrm{ET}}=&\sqrt{\frac{1}{N}\sum_1^N(\theta_{est}-\theta_{exp})^2}\\
R^2_{\mathrm{ET}}=&1-\frac{SSR}{SST}=1-\frac{(\theta_{exp}-\overline\theta_{exp})^2}{(\theta_{est}-\theta_{exp})^2}\\
\overline{\eta}=&\frac{1}{N}\sum_1^N\frac{\theta_{est}-\theta_{exp}}{\theta_{exp}}
\end{align}
where the overline represents arithmetic mean, $RMSE$ is root-mean-square error, RMSE and R$^2$ subscripts (ET) denote error types, $\theta_{exp}$ and $\theta_{est}$ are experimental and estimated values for dynamic contact angle and $N$ is the number of data points.

A summary of the models, best fits and goodness of fit indicators can be found in table~\ref{fit}, while figure~\ref{graph1} shows the best fits obtained by the three models against experimental data. All of the three models proved capable of predicting contact angle, although with varying success, for all regions of capillary number and all static contact angles. But as can be seen in table~\ref{fit} the hydrodynamic model, as a universal model, outperforms the other two by most measures. As expected~\ref{graph1} clearly shows that this model is accurate at higher capillary numbers but the data becomes increasingly scattered at lower Ca and while the same thing happens in all models it is particularly emphasized in this model.

Interestingly, while the molecular-kinetic model works just as well as the hydrodynamic model when used for a particular solid-liquid-gas system~\cite{seveno2009dynamics}, it has a poor performance as a universal model. Given that the model is fitted for velocity, it seems likely that molecular-kinetic model is sensitive to solid-liquid-gas system variations.

Furthermore, given the source data we could not calculate measurement uncertainties and average absolute error for the hydrodynamic model only shows marginal improvement over the empirical model, inviting suspicion of overlap in error ranges. But, average relative error, R$^2$ and RMSE values are much better for the hydrodynamic model. This model is very competent in predicting dynamic contact angles for every range of capillary number and static contact angle included in this study.

\begin{table*}
   \centering
    \caption{A summary of the models and goodness of fit indicators}
    \label{fit}
    \renewcommand{\arraystretch}{1.05}
   \begin{tabular}{ lccc }
    \hline
     & Empirical & Molecular-kinetic & Hydrodynamic\\ \hline 
      Equations & $H=a\textrm{Ca}^b$ & 
     $ \sigma(\theta^{2}_{\textrm{D}}-\theta^{2}_{\textrm{S}})= a u^{b}$ &
     $\theta^{3}_{\textrm{D}}-\theta^{3}_{\textrm{S}}=a \textrm{Ca}^{b}$ \\\hline
     
     \multicolumn{4}{l}{General ($10^{-16}>\textrm{ln(Ca)}>10^{-2}$)}\\\hline
     Constants & \pbox{20cm}{$a=2.641$\\ $b=0.5428$}& \pbox{20cm}{$a=1.261$ \\ $b=0.7918 $} & \pbox{20cm}{$a=84.67$ \\ $b=0.9116$}\smallskip\\
     $\overline{\epsilon}$ (rad) & 0.1156 & 0.2623 &  0.1012\\ 
    RMSE & 0.1217 & 0.3089 & 0.0921\\
   $ \overline{\eta}$ & 0.1138 & 0.4450 & 0.1501 \\ 
   R$^2$ & 0.8001 & 0.8438 & 0.9068 \\  \hline
    
    \multicolumn{4}{l}{High capillary region ($\textrm{Ca}>10^{-4}$)}\\\hline
    
     Constants & \pbox{20cm}{$a=2.976 $\\ $b=0.5808$}& \pbox{20cm}{$a=1.351$ \\ $b=0.8179 $} & \pbox{20cm}{$a=83.71$ \\ $b=0.906$}\smallskip\\
    $ \overline{\epsilon}$ (rad) & 0.3625 & 0.3008 &  0.1005\\ 
   RMSE & 0..2977 & 0.2977 & 0.0911\\
    $\overline{\eta} $ &0.2230 & 0.4197 & 0.1214 \\ 
   R$^2$  & 0.8874 & 0.8530 & 0.8932\\  \hline
    
    \multicolumn{4}{l}{Low capillary region  ($\textrm{Ca}<10^{-4}$)}\\\hline
    
             Constants & \pbox{20cm}{$a=1.937 $\\ $b=0.4632$}& \pbox{20cm}{$a=0.8506$ \\ $b=0.684 $} & \pbox{20cm}{$a=58.41$ \\ $b=0.8169 $}\smallskip\\
     $\overline{\epsilon}$ (rad) & 0.1151 & 0.1426 &  0.1034\\ 
  RMSE & 0.0784 & 0.1578 & 0.0880\\
    $\overline{\eta}$  &0.0743 &0.4776 & 0.2265 \\ 
  R$^2$ & 0.6371 & 0.8056 & 0.9651 \\  \hline
    \end{tabular}
\end{table*}

\input{graph1}

Given that hydrodynamic forces become less dominant at lower capillary numbers and give way to molecular forces, and the sharp contrast evident in accuracy of the models in low and high capillary numbers in~\ref{graph1}, it is necessary to compare the models' capabilities at different capillary number regions as well. To this end we divided the results into two sections, capillary numbers higher and lower than $10^{-4}$, and repeated the fitting procedure for each section. The results for each region are also shown in table~\ref{fit}. Due to small contact angle changes in low capillary region they have minimal effect on general models, therefore high capillary region models are almost identical to the general models which might limit the benefit of making such distinctions since the goal of the present models is to simplify dynamic contact angle estimation. 

The hydrodynamic model is actually less accurate at low capillary numbers than it is as a universal model, meaning that as expected it is more accurate at higher capillary numbers where hydrodynamic forces are dominant. Surprisingly, despite significantly improved accuracy, the molecular-kinetic model's predictions are still less accurate than the other two models which means that this model is unsuitable for generalizations and simplifications even in lower capillary numbers. The empirical model shows much better accuracy at low capillary numbers than overall and is the most accurate model in this region by most measures.

For a simplified model to be useful it should also be capable to predict values in each experimental setup and not only the over all data. Table~\ref{breakdown} shows RMSE and average absolute error of general models (first section of table~\ref{fit}) for each experimental setup included in the current study. As previously mentioned, the experiments were intentionally chosen to represent a wide range of conditions and include high and low capillary numbers, non-Newtonian fluids and various flow geometries and scales. As can be seen, the hydrodynamic model is performing slightly better than the empirical model once again and shows consistent predictive capabilities in the diverse experimental conditions included in the present study, while the empirical model is a close second and MKT shows limited potential as a simple and universal model.
\begin{table*}
   \centering
    \caption{A summary of goodness of fit indicators for each experimental study}
    \label{breakdown}
    \renewcommand{\arraystretch}{1.05}
   \begin{tabular}{ lccccccc }
    \hline
     Study & \multicolumn{2}{c}{Hydrodynamic}&\multicolumn{2}{c}{MKT}&\multicolumn{2}{c}{Empirical}\\\hline
      &\centering{RMSE}& $\overline{\epsilon}$&\centering{RMSE} & $ \overline{\epsilon}$ & RMSE & $ \overline{\epsilon}$ \\ \hline 
     Hoffman et al.~\cite{hoffman1983study} & 0.0859 & 0.1027 &  0.0598 & 0.5800&  0.0543 & 0.0638\\ 
    Seebergh et al.~\cite{seebergh1992dynamic}\\
     \hspace{3.4mm} High Ca& 0.0210 & 0.0523 & 0.0530 & 0.1423&  0.0408 & 0.0377\\
    \hspace{3.4mm}  Low Ca& 0.0438 & 0.0490 & 0.1561 & 0.5145&  0.0142 & 0.1782\\
    Kim et al.~\cite{kim2015dynamic} & 0.0369& 0.1569 & 0.0975 & 0.4549&  0.0877 & 0.1523\\ 
    Wang et al.~\cite{wang2007spreading} & 0.0389 & 0.0559 & 0.2960 & 0.5234&  0.0467 & 0.1168\\
    Schneemilch et al.~\cite{schneemilch1998dynamic} &0.0604 & 0.0522 & 0.0431 &0.3234 &  0.0616 & 0.0918\\ 
    Rio et al.~\cite{rio2005boundary} &0.0532  & 0.0581 & 0.2079 & 0.1128&  0.0284 & 0.0565\\ 
    Seveno et al.~\cite{seveno2009dynamics}\\ 
      \hspace{3.4mm} DBP on PET&0.0856 & 0.0735 & 0.0388 & 0.0552&  0.0599 & 0.1166\\ 
 \hspace{3.4mm}  PDMSOH1&0.1269 & 0.1055 & 0.1815 & 0.1128&  0.2439 & 0.2694\\
 \hspace{3.4mm}   Squalane& 0.0250 & 0.0379 & 0.0394 & 0.0583&  0.0338 & 0.0887\\
 \hspace{3.4mm}  PDMSOH1& 0.0211 & 0.0380 & 0.0446 & 0.0586&  0.0459 & 0.0600\\
     \hline  
    \end{tabular}
\end{table*}
Even though all of these models, and particularly the hydrodynamic inspired model, are promising and capable of predicting dynamic contact angle for all of the available experiments, further experimental investigations are required. For example, we have limited understanding of the wetting line and dynamic contact angle behavior in extremely low capillary numbers. The slip and stick behavior of the wetting line and the resulting data scattering in low capillary numbers demand that we have a higher density of experimental data in this region.

\section{Summary and Conclusions}
In this work we have revisited the dynamic contact angle prediction problem, which has been surprisingly dormant for quite some time. Noticing the need for a simple, accurate and more comprehensive model, we gathered a sizable data pool from various studies and derived simple intuitive forms from the hydrodynamic and molecular-kinetic models. These models, along with an empirical model, were fitted to the data pool and compared with each other. The model inspired by the hydrodynamic model performed better than other models by every statistical measure and was relatively accurate at all capillary numbers and static contact angles. Then, in order to compare accuracy of the models at various capillary numbers we separated the data into two sections, capillary numbers higher and lower than $10^{-4}$, and repeated the fitting process for each section. The models obtained at high capillary numbers was almost exactly the same as the general models but the results showed that the empirical model slightly edges out the hydrodynamic model to give the most accurate results at lower capillary numbers. 

\FloatBarrier

\bibliographystyle{elsarticle-num}
\bibliography{DCA}

\end{document}